# Extending Standard Atomic Kernel Model
# with New Interpretation of Strong Forces

Dr Ing Hongguang Yang, Dr ret nat Weidong Yang

## Introduction

This paper gives an extended model of the atomic nucleus - we call it the **YY model**, which allows a new description for strong forces from the well-known Standard Model. The forces that hold the nucleus together (protons and neutrons) can be expressed in a new way. Based on the YY model, more structural description details for an atomic nucleus will be possible than is the case with the conventional description. The YY model is compatible with the standard model. However, it allows for very delicate considerations when modelling the distribution of protons and neutrons within or around an atomic nucleus. Furthermore, it can explain many subatomic processes in an elegant way. The YY model predicts some subatomic aspects that can be explored in a deeper step. Also a native common root for the description of macrophysics (space, dark matter and cosmology) is given.

In the new model, the so-called "pairing space link PSL" and "triple space link TSL" are introduced as mechanisms to describe subatomic constructions and transformations. "Yin" and "Yang", which represent conceptual units of one third of electric charge, are defined and used to control nuclear transformation rules by preserving their numbers. In addition, a new "**Y particle**" is postulated in connection with TSL, which plays a central role in the interpretation of nuclear bonds, electron-positron annihilations, and the cosmological configuration of space and the energy-matter life cycle.

Following a model-driven approach - a methodology widely used in computer science and informatics - the investigations in this paper will completely dispense with mathematical formulations. In addition, energy balances in the transformations are left out. Many different aspects that can be derived from this new mechanism can then be verified. We hope to gain the confidence of many other physicists in the YY model. They could make detailed theoretical and experimental verifications by adding additional descriptive elements and linking the YY model to the known quantum field theories and the widely accepted big bang theories for space and cosmology, especially for research on dark matter.

## 1.  Alternative Up and Down Quark Description

The starting point is the conventional description of the elementary particles, up quark and down quark, in considering their essential properties: An up quark possesses 2/3 positive of an electrical charge, whereas a down quark has 1/3 negative of an electrical charge (Ref [1]). Consequently, two basic building units are elaborated as following:

> ➢ Unit corresponding a third of negative charge, symbolized by " **-** ", called by "**yin**"
> ➢ Unit corresponding a third of positive charge, symbolized by " **+** ", called by "**yang**"

These two units, supposed its physical existence, attract each other, and they keep from each other too, so that there is a stable state of the "pairing" of the both units, building a space link, called by "**pairing space link PSL**":

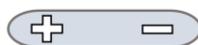

Figure 1: Yin – Yang paring for building a physical space link

This construction can be expressed by {- <> +}. As will be seen later, a space link by pairing is a part of the basic concept of YY model for materialization from space and energy, which means to assign the pairing (space link) a matter and a space properties (mass and dimension). The curly bracket "{ }" symbolizes this fact.

By using PSL, quarks will be (re-)constructed as following

- A down quark contains a yin-unit, expressed as symbol (-)
- An up quark contains two yang-units and a space link, expressed by (++{+ | - })

Drawing out the geometric forms of up quark and down quark turns Figure 2:

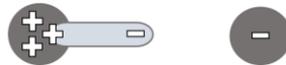

Figure 2: Geometric Forms of Up Quark and Down Quark

Parenthesis "( )" in expressions and shadows in Figures symbolize the mass properties of the quarks.

## 2. Re-Modeling of Neutron and Proton

Because a neutron is made-up of two down quarks and one up quark and, a proton is made-up of one down quark and two up quarks (Ref [1]), the YY model turns out the Figure 3.

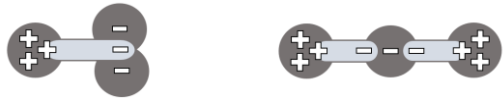

Figure 3: Re-Model of Neutron and Proton

This re-modeling of neutron and proton reveals more details of their inner structure.

The net summation of the positive and negative electrical charges, respectively for a proton and for a neutron, correspond to the classic description. The only difference comes from the space links: they hold the nucleus together.

The following expressions are equivalent descriptions for neutron and proton as in Figure 3:
Neutron: (++{+ | -})(-)(-)     Proton: (++{+ | -})(-)({ - | +}++)

The introduced meta model for atomic kernel here can already make concrete predictions for the interaction behaviors of neutron and protons, with each other or with electrical fields. For example:

- ✓ A neutron has a space orientation respective to the inner charge distribution, namely the axis from the positive to negative pool, as showed in the Figure 4(a), though the whole system of electrically neutral. From this point of view, a neutron, released in an electrical field, should orient its positive end towards the negative side of the electrical field, Figure 5(a).

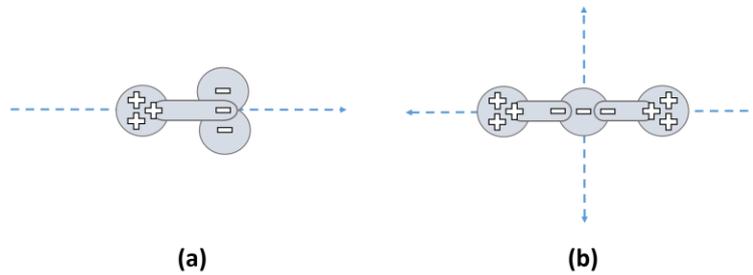

Figure 4: Space Rotation Axes of Neutron and Proton

- ✓ A proton has a rotation symmetry respective to the axis through the charged poles ("positive – negative – positive"). The axis orientation from left to right does not differ from right to left, Figure 4(b). From this point of view, a proton, released in an electrical field, should orient its symmetry axis perpendicularly to the electrical field, Figure 5(b). A possible electrostatic deformation of the proton structure is also indicated there.

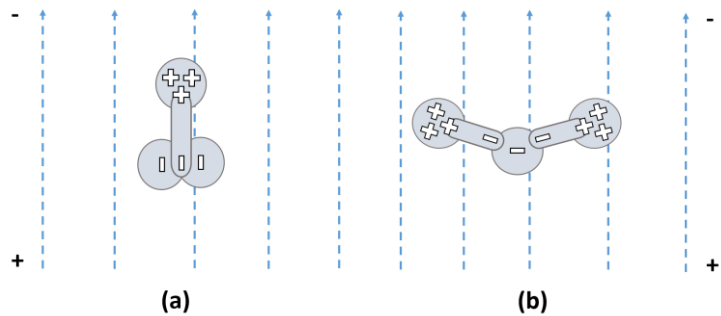

Figure 5: Orientation of Neutron and Proton in an Electrical Field

By given the proton model, as described above, further descriptions for atomic kernels of the hydrogen isotopes deuterium (H-2, Ref.[3]) and tritium (H-3, Ref.[3]) will continue.

## 3. Triple Space Link and its Role for the Atomic Kernel

To model complex kernels, a new construct "**triple space link** (**TSL**)" is necessary:

- ➢ Three PSLs can bind themselves together on their yang units to build a bounded aggregate. Their common node corresponds a positive electrical charge, see Figure 6(a).
- ➢ Similarly, three PSLs can bind themselves together on their yin units to build a bounded aggregate. Their common node corresponds a negative electrical charge, see Figure 6(b).

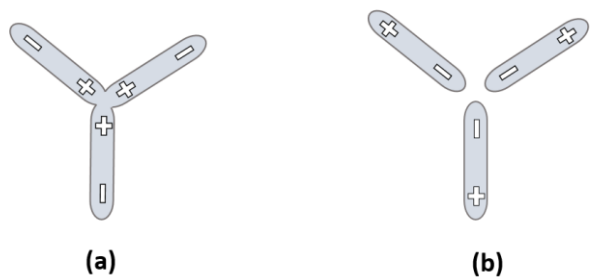

Figure 6: Triple Space Links, Positive and Negative Charged

It is still unclear, how three basic units are linked in a charged node from physical perspective of view.

Furthermore, see also the difference between Figures (a) and (b) in the node display:

- ✓ There is an asymmetry between positive charged TSL and negative charged TSL: Whereas a positive charged TSL is in YY model is pre-created and bound strongly, a negative charged TSL is a mixture of yin pools from positive charged TSLs, of up quarks or of a down quarks. Figure 7 gives all possible combinations for negative charged TSLs, inclusive neutron and proton.

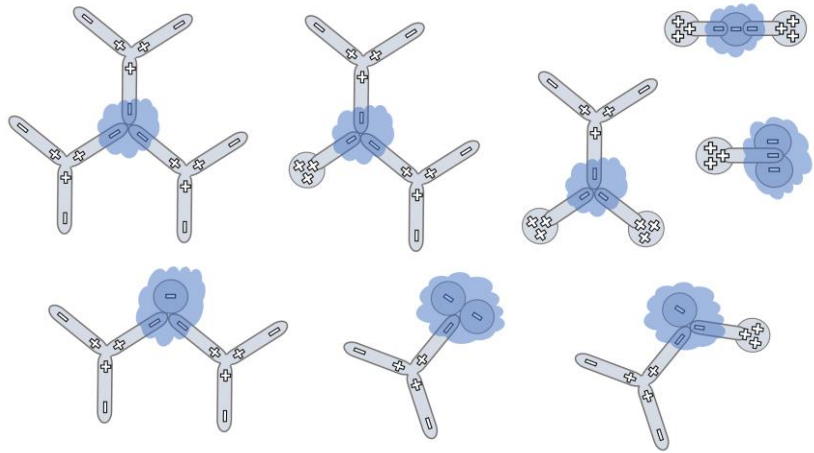

Figure 7: All Possible Combinations for Negative Charged TSLs

This asymmetry is corresponding the asymmetry between down and up quarks, which will be discussed in a very later section in concerning the origins of cosmology. Following rules apply:

- ✓ Pairing space link (PSL) can be embedded in an up quark, as described in section 1
- ✓ Triple pairing space links can be merged on the common positive ends (yang-ends) to build a triple space link (TSL), Figure 6 and 7;
- ✓ Atomic kernel building is a combination of up quarks by their yin-ends, TSLs by their yin-ends, and down quarks by themselves, section 4 and more;
- ✓ **Conservation of yin's and yang's**: In all transformation processes, the whole number of yin's and the whole number of yang's remain unchanged.

## 4. Hydrogen Atom Kernel H-2 and Re-Interpretation for Strong Force

The starting point is formally going to disassemble an existing proton and an existing neutron and get all their building quarks, see Figure 8. Parts are colored for easy identifying before and after the transformation.

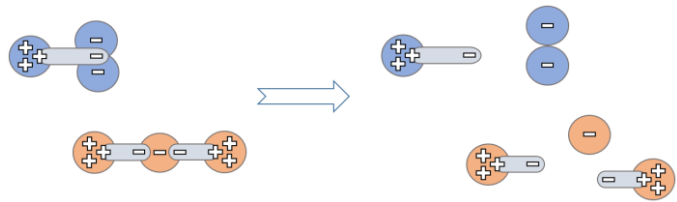

Figure 8 Disassembling a Proton and a Neutron to their Building Quarks

The next step is to reassemble them to a single atomic kernel for building a deuterium (Ref. [3]), with a positive electrical charge unit and a mass unit of two (namely the mass summation of a proton and of a neutron). There are many possible ways for reassembling the given quarks, turning out different (quark-) configurations. Only certain configurations of down and up quarks are able to satisfy the charge and mass restrictions for a deuterium.

Starting from rules for quark configurations (at the end of section 3), more rules will be introduced in later section, while considered atomic kernels become more complex. In the case for reassembling a deuterium, a new triple space link TSL is needed, which combines all given down and up quarks for getting a consistent deuterium atomic kernel, as described in Figure 9.

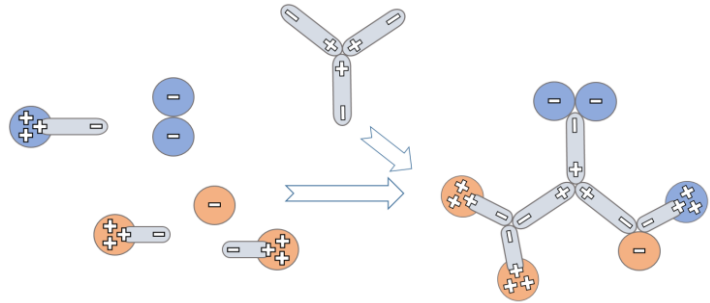

Figure 9: Reassembling the Given Quarks to a Deuterium by Adding a TSL

The upcoming of a TSL in the process of reassembling is a very important point and will be treated in further discussions later.

In consideration of the electrical charges of the whole aggregate above: The net summation of the positive electrical charges is three time 2/3 (=> two positive charge units) and the net summation of the negative electrical charges is three time 1/3 (=> one negative). The net balance of all electrical charges turns out one positive.

In consideration of the mass of the whole aggregate above: The net summation of all three up and all three down quarks results a complete mass of two atomic units (three times a third from up quarks, plus three time a third from down-quarks (3x{1/3}+3x{1/3} = 6/3 = 2).

The one positive triple link and one negative triple link in the interior of the aggregate are equilibrated in charges.

The strong forces of standard model holding the proton and the neutron together are re-interpreted by links of different sorts, without using dynamic colored charges of quarks, anti-quarks and gluons (Ref. [2]).

Note please also the drawing of the atomic aggregate in a two-dimensional plane: For a three-dimensional description in real world, electrostatic attractions and repulsions of up and down quarks around the whole aggregate must be taken into account. In the case of Figure 9, the double down quarks on the top (or/and the double up quarks on the left) must be arranged perpendicular to the plane for balancing electrostatic forces.

## 5. Hydrogen Atom Kernel H-3

By applying the reassembling schema in the section above, it is easily get a consistent atomic kernel model for the tritium (Ref. [3]). Firstly, an existing deuterium and an additional neutron are disassembled in Figure 10.

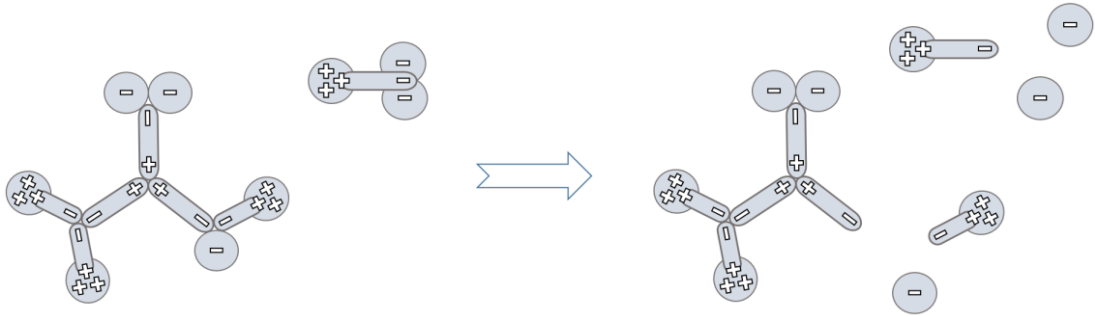

Figure 10: Disassembling a Deuterium and a Neutron

Thereafter, all parts are reassembled together, by adding a new triple space link TSL, resulted as a tritium kernel in Figure 11(a). It is also possible to get a reassembled alternatively model in Figure 11(b).

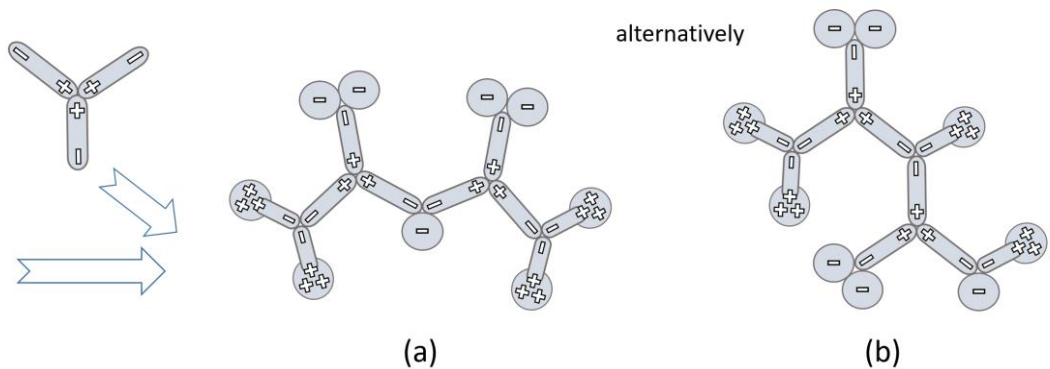

Figure 11: Two Alternative Tritium Kernel Models Reassembled by Adding a TSL

Please also note the upcoming of a TSL in the process of reassembling.

In consideration of the electrical charges in the aggregate above: The net summation of the positive electrical charges is 4x{2/3} (four up quarks times 2/3) and the net summation of the negative electrical charges is 5x{1/3} (five down quarks times 1/3). The net balance of all electrical charges turns out one positive (4x{2/3}-5x{1/3} = 3/3 = 1) .

In consideration of the mass in the aggregate above: The net summation of all four up and all five down quarks results a complete mass of three atomic units (four times a third from up quarks, plus five time a third from down-quarks 4x{1/3}+5x{1/3} = 9/3 = 3).

The two positive triple nodes and two negative triple nodes in the interior of the aggregate are equilibrated in charges.

Similarly, the classic strong forces holding the proton and the neutrons are re-interpreted by links of different sorts. This is the same mechanism for holding one proton and one neutron together.

## 6. Helium Kernel He-4 and its Isotope He-3

This section will describe the helium atomic kernel He-4 in YY model, constructed by two protons and two neutrons. Furthermore, one of its isotopes He-3 will also be considered, with two protons plus one neutron.

In the case of He-4, a formal reassembling from two H-2 atom kernels is done. Figure 12 firstly disassembles two H-2 atom kernels respectively.

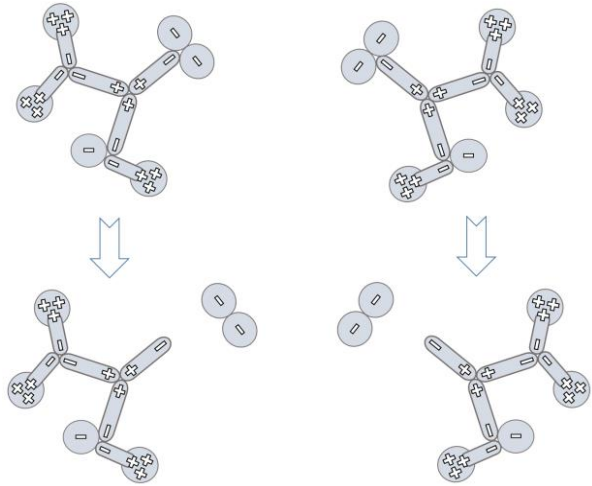

Figure 12: Disassembling two Hydrogen Atomic Kernels H-2

By adding one triple space link TSL and reassembling them all, a stable helium kernel He-4 is resulted, see Figure 13.

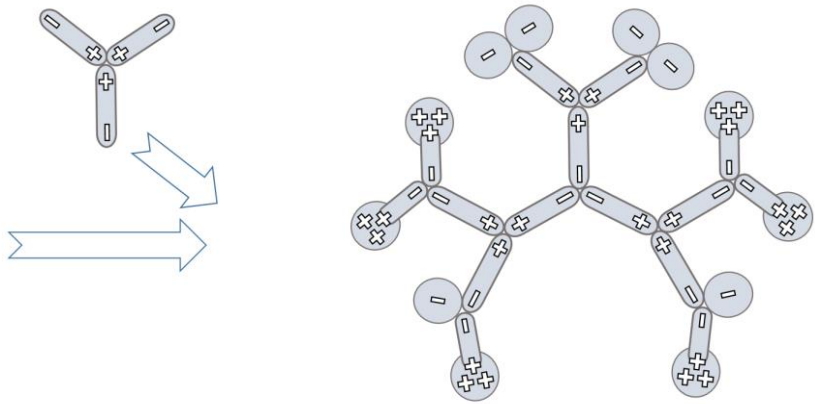

Figure 13: Reassembled Atomic Kernel for a Helium-4 from Figure 12 by Adding a TSL

Consideration of the electrical charges: The net summation of the positive electrical charges is 6x{2/3} (six up quarks times 2/3) and the net summation of the negative electrical charges is 6x{1/3} (six down quarks times 1/3). The net balance of all electrical charges turns out two positive (6x{2/3}-6x{1/3} = 6/3 = 2) .

Consideration of the mass: The net summation of all six up and all six down quarks results a complete mass of four atomic units (six times a third from up quarks, plus six time a third from down quarks 6x{1/3}+6x{1/3} = 12/3 = 4).

The three positive triple nodes and three negative triple nodes in the interior of the aggregate are equilibrated in charges.

The same mechanism is used for holding up and down quarks together makes a re-interpretation for the strong forces.

As next step, atomic kernel of the helium isotope He-3 is considered, firstly by disassembling a hydrogen H-1 and a deuterium H-2 kernel respectively, as in Figure 14.

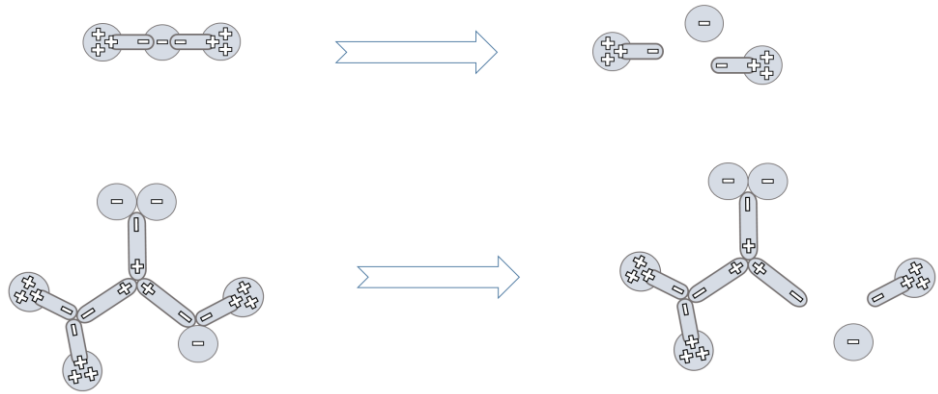

Figure 14: Disassembling Hydrogen H-1 and Hydrogen H-2 Kernel

By adding one triple space link TSL and reassembling them all, the result helium kernel He-3 is resulted in Figure 15.

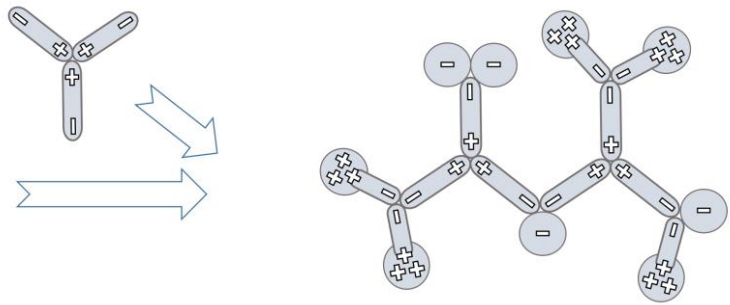

Figure 15: Reassembling Atomic Model for a Helium Isotope He-3 by Adding a TSL

Consideration of the electrical charges: The net summation of the positive electrical charges is 5x{2/3} (five up quarks times 2/3) and the net summation of the negative electrical charges is 4x{1/3} (four down quarks times 1/3). The net balance of all electrical charges turns out two positive (5x{2/3}-4x{1/3} = 6/3 = 2) .

Consideration of the mass: The net summation of all five up and all four down-quarks results a complete mass of three atomic units (five times a third from up-quarks, plus four time a third from down-quarks 5x{1/3}+4x{1/3} = 9/3 = 3).

The two positive triple nodes and two negative triple nodes in the interior of the aggregate are equilibrated in charges.

## 7. Alternative Kernel Configurations for Helium He-4

YY model makes it possible that multiple consistent atomic kernel models (alternative configurations) for helium He-4 exist. One of the alternatives to Figure 13 is to reassemble Figure 12, by adding two triple space links, to the Figure 16.

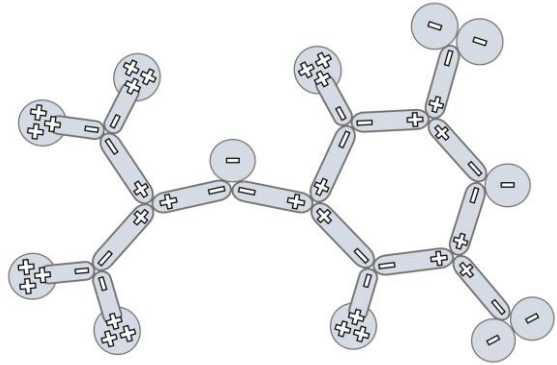

Figure 16: A Possible Alternative Atomic Kernel Model for Helium He-4

Another alternative is, also by adding two triple space links, to reassemble Figure 12 to Figure 17.

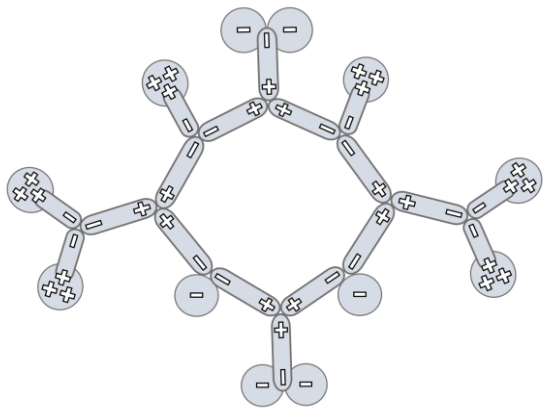

Figure 17: Another Possible Alternative Atomic Kernel Model for Helium He-4

All common things of models in Figure 13, 16 and 17 for helium atom kernel He-4 are

- The numbers of up quarks (6) and the numbers of down quarks (6)
- The electrical charge balance (positive 2), and
- The mass summation units of all quarks (3)

The internal triple space links vary from alternatives: Figure 13 possesses three positive TSLs and their balancing negative nodes; Figure 16 and 17 possess four positive TSLs and four balancing negative nodes respectively. Generally as a configuration rule:

- The number of positive TSLs is equal to the number of negative balancing nodes (**Internal Charge Balance ICB**)

ICB is important for the stability of aggregate structures, though the governing mechanism must be still investigated mathematically and physically.

Moreover, a TSL must possess a mass unit that is essential smaller in comparing to the mass of an up or down quark – these are in the order of a third atomic mass unit.

## 8. Process for Generating Electron and Positron

This section discussed how the YY model understands electron and positron, their generation processes, by using prototype for electron and, by splitting up quarks for positron.

For generating an electron, YY model considers formally a nuclear interaction of two neutrons with each other (disassemble them and reassemble the parts). One can get a proton and three "freed" down quarks that build together a **prototype** for a negative charged particle, successively decayed to an electron and neutrino/antineutrino of different generations, as Figure 18 demonstrates.

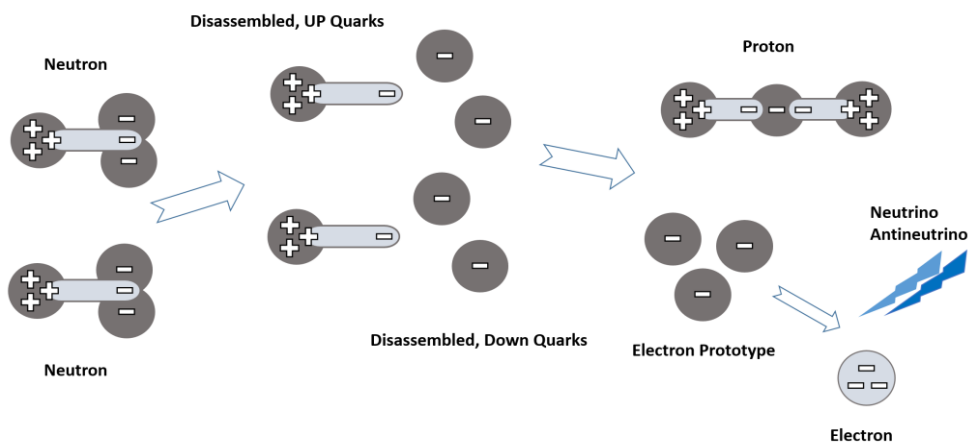

Figure 18: Generating Process for Electron

Essential results for generating electron are:

- ➢ An electron in the YY model is a deep-bound yin's, originated from three down quarks (likely a positive triple space link is a deep-bound yang's from three paring links)
- ➢ Kernel transformation model:
  **2 n ➔ p$^+$ + electron prototype ➔ p$^+$ + e$^−$ + neutrino, antineutrino**

The main difference between YY model and the well-known processes in standard model like muon decay ($\mu^- \to e^- + \bar{\nu}_e + \nu_\mu$) (Ref. [3], [4]) and beta decay (n ➔ p$^+$ + e$^−$ + $\bar{\nu}_e$) (Ref. [3], [4]) is the involved number of neutrons. Furthermore, the electron origination (prototype) and resulted properties (deep-bound three yin's) differ.

Generating positron is interpreted by the "inverse beta decay" (Ref. [3]): $\bar{\nu}_e + p \to e^+ + n$, as displayed in Figure 19.

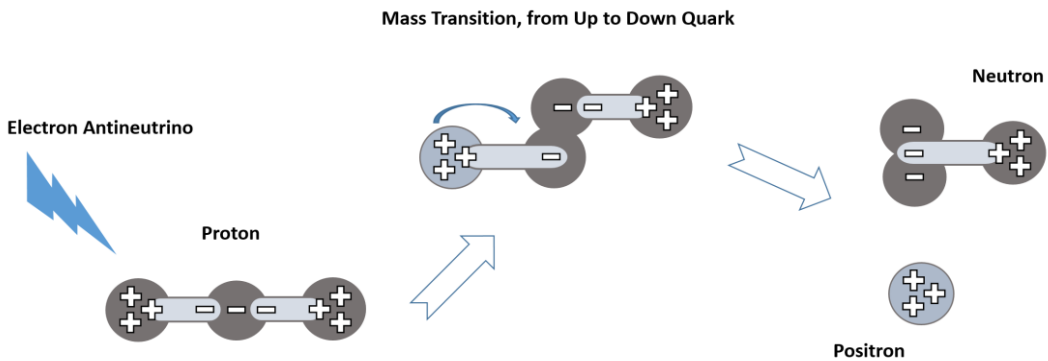

Figure 19: Generating Process for Positron

Obviously, the antineutrino strikes one up quark, shifted the mass (grey shadowed) from its positive charge pool to its negative pool becoming a down quark to combine the rest of the aggregate to a neutron, while the positive part of the concerned up quark becomes a positron.

Of course, the widely accepted standard particle model postulates that a free electron does not possess a measurable dimension and, that an electron is not dividable to smaller elementary particles.

From the perspective of YY model, electron as deep-bound yin's can still have a smaller dimension unit than our current physical detective ability. Furthermore, electron as "no dividable particle" is also a fact, with one difference:

The only way to reassemble an electron is its collision with a positron that leads to annihilation of the both particles and emission of gamma radiations. In the YY model, this collision will create a reassembled triple space link TSL (see next section) which is "given back" to the space and "becomes" a part of matter.

## 9. Electron-Positron Annihilation in YY Model, TSL Existence and Y Particle

The YY model conserves the participating yin's and yang's in all transformation processes. The electron-positron annihilation described in standard model (Ref. [3], [5]) $e^+ + e^- \rightarrow 2\gamma$ will be extended in YY model by generating a TSL obeying this conservation, see Figure 20.

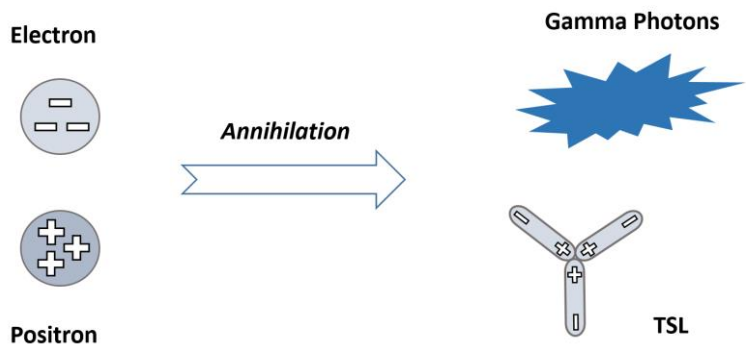

Figure 20: Electron-Positron Annihilation by Generating Gamma Photons and a TSL

TSL is a prediction of YY model. It can be imagined as ashes, after wooden has burned out. Every TSL possesses a mass – defined as "**TSL Mass Unit**", which must be much smaller than the mass of electron and positron, but still not negligible. It could be possible to detect it indirectly by using more precise energy-matter balance between predictions and observations, also to estimate the absolute mount of TSL mass unit.

Actually, the electron-positron annihilation builds a part of energy-matter lifecycle, originating from TSLs at very beginning of cosmology. The third to the last section will give a more detailed description for this lifecycle. In addition, the TSL origination builds the basis for the antimatter systematics of the YY model (out of the scope of this paper).

We call TSL as **Y particle**, in order to motivate physicians to verify its existence.

## 10. Rules for Electrical Charge Units and Mass Units of Atomic Kernel

The early modelling approaches in this paper have showed, that the net electrical charges of an atomic kernel turn out from the summation of all building outer vertexes – all up and down quarks combinations. For clearness, some more terminologies are introduced here (see also Figure 21):

- **Neutronhead**: Triple end node containing two down quarks – possesses one negative electrical charge unit and 2/3 atomic mass unit;
- **Protonhead**: Triple end node containing one down quark – possesses one negative electrical charge unit and 1/3 atomic mass unit;
- **Protonid**: Single end node containing one up quark – possesses one positive electrical charge unit and 1/3 atomic mass unit.

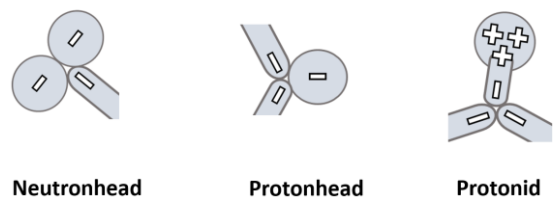

Figure 21: Neutronhead, Protonhead and Protonid

Additionally to the properties for electrical charges and mass units described above, there are also rules for balancing the numbers of protons and neutrons in an atomic kernel:

- Every protonhead needs two corresponding protonids for building a proton within an atomic kernel aggregate (number of protonheads = number of protons);
- Every neutronhead needs one corresponding protonid for building a neutron within an atomic kernel aggregate (number of neutronheads = number of neutrons);

Thus, the occurrences of protonids within an atomic kernel aggregate are not determining factor for calculating the numbers of protons and neutrons: They are only for satisfying the other requirements from the occurrences of protonheads and neutronheads.

This implies another asymmetry between up quark and down quark: The occurrences of down quarks (single or in pair) are essential for determining the protons and neutrons assembled in an atomic kernel.

Furthermore, a special rule has to be obeyed by the whole aggregate:

- The number of positive triple space links (TSLs) and the number of negative binding vertexes must balance in total (=> **Internal Charge Balance ICB**), as already discussed in the early Figures (9, 11, 13, 16 and 17).

In following Figure of the He-4 atomic kernel, colored texts annotate the internal charge balance (blue), respectively the external charge balance (red). "Internal" means pairing space links (PSLs) embedded within triple space links (TSLs). "External" means all surrounding protonheads, neutronheads and protonids.

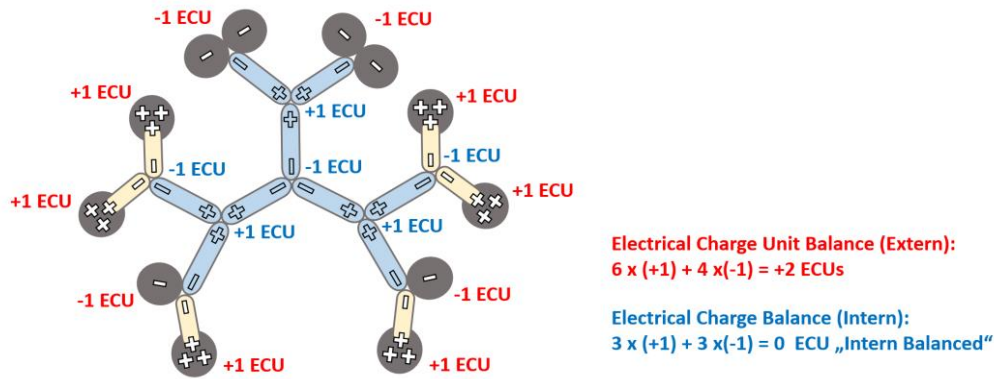

Figure 23: Electrical Charge Balances of Helium Kernel He-4

The internal charge balance (ICB) guarantees the electrical neutrality of the binding element structure (PSLs, TSLs) so that the whole electrical charge balance of an atomic kernel just results from the summation of all surrounding protonheads, neutronheads and protonids.

All these rules give us the guidelines for assembling and reassembling new atomic kernels by disassembling the original atomic kernels, in a consistent way.

The atomic mass unit calculation for a kernel aggregate is simple and straightforward:

- The atomic mass units of an atomic kernel results from the summation of all atomic mass units of surrounding protonheads, neutronheads and protonids, divided by three;
- The TSL mass units results from the number of existing TSLs.

In following Figure of the He-4 atomic kernel, colored texts annotate atomic mass units of all protonheads, neutronheads and protonids (green), respectively the TSL mass units (blue).

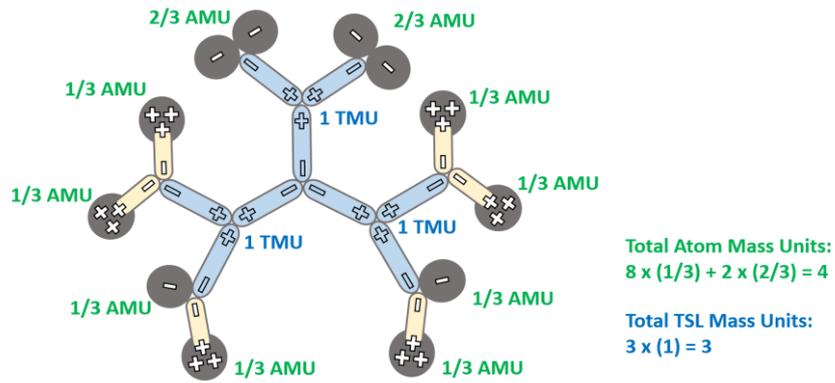

Figure 24: Atomic Mass Units and TSL Mass Units of Helium Kernel He-4

## 11. Stellar Nuclear Fusion Process

In this section, some major parts of stellar fusion reactions (Ref. [6], [7]) will be described – separated in four subsections. Mechanisms of YY model discussed in early sections will be applied here for understanding the internal structure-related changes in the fusion reactions, without considering energy transitions details.

### 11.1. Nuclear Fusion to Deuterium H-2

The beginning stage of stellar nuclear fusion produces a hydrogen isotope H-2 by merging two protons H-1 and generating a positron and neutrino: H-1 + H-1 ➔ H-2 + $e^+$ + υ.

Following the same construction procedure by disassembling the both H-1 and reassembling their elements to get H-2 and co. To ensure the model consistence by obeying the rules (section 10), a new triple space link TSL is needed (Figure 25).

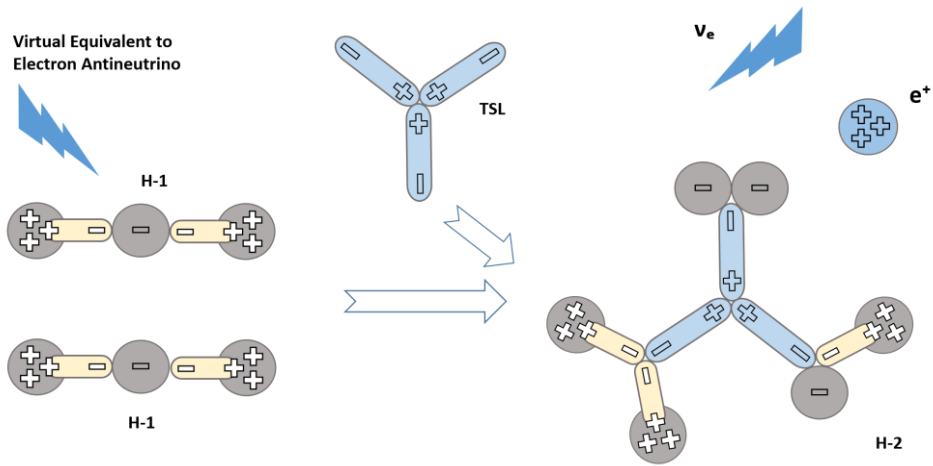

Figure 25: Nuclear Fusion to H-2

The complete transformation can be seen in two steps. In the first step, a proton is virtually transformed into a neutron by an inverse beta decay (Figure 19) and, in a further step, the virtual neutron and a second proton are merges into a deuterium (Figure 9) by adding a TSL.

The interpretation of the first step is equivalent to assuming the transition of an up quark to a down quark, a positron and an electron neutrino (u => d + e$^+$ + $\upsilon_e$), Figure 26.

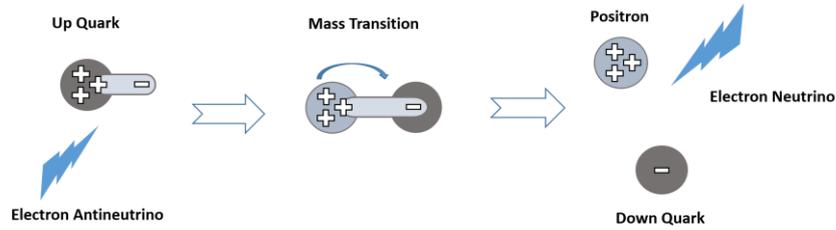

Figure 26: Transition of an Up Quark to Down Quark, Positron and Neutrino

However, what happens to the excessive positron? Moreover, where does TSL come from? Obviously, the both questions have something strongly to do with each other. This will be explained in the section 11.4, "Mechanism for Producing TSLs in Stellar Nuclear Fusion".

## 11.2. Nuclear Fusion to Helium Isotope He-3

The next stage of stellar nuclear fusion produces a helium isotope He-3 by merging a proton H-1 and a hydrogen isotope H-2, emerging gamma photon: H-1 + H-2 ➔ He-3 + γ (Figure 27). Following the YY model, a new TSL is needed, accompanying the gamma emerging.

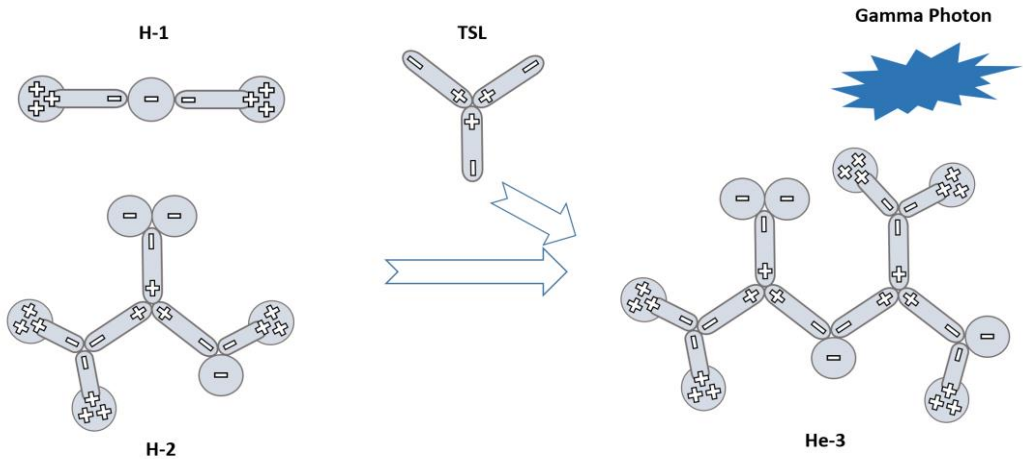

Figure 27: Nuclear Fusion to He-3

## 11.3. Nuclear Fusion to Helium He-4

In a further stage of stellar nuclear fusion, a stable helium He-4 and two protons H-1 will be produced by merging two helium isotopes He-3: He-3 + He-3 ➔ He-4 + 2 p + 12.9 MeV.

The structure of helium He-3 is already described in early section (Figure 15). The fusion process after YY model is described in Figure 28. To illustrate clearly, and for better identifying, parts are colored in aggregates before and after the fusion,

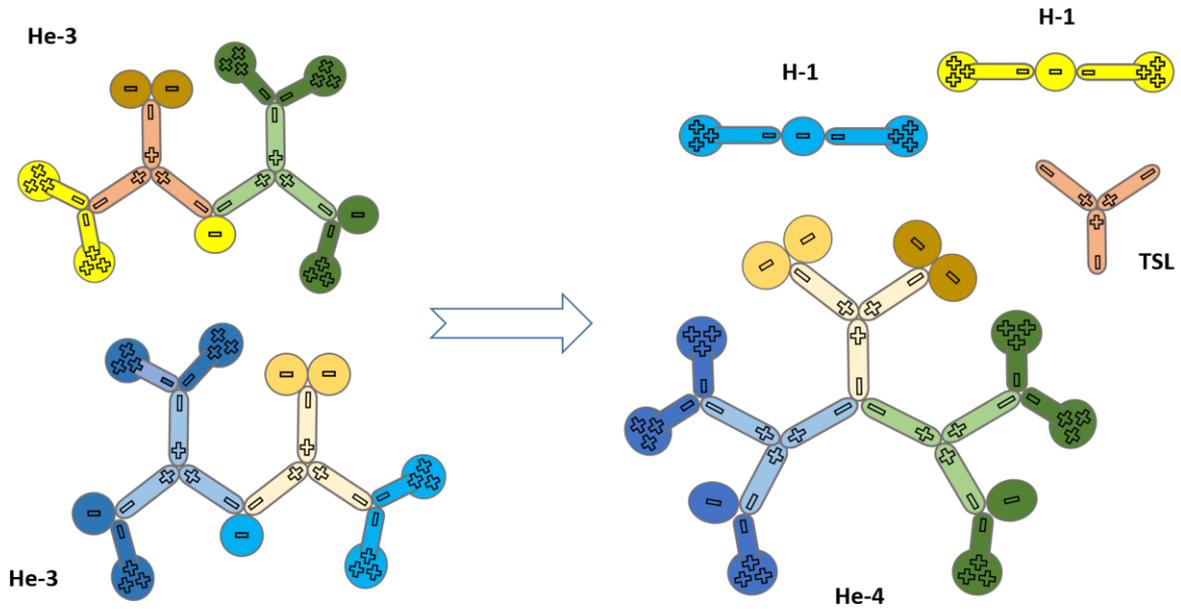

Figure 28: Nuclear Fusion to He-4

Be aware that this fusion stage releases an existing TSL, which can be re-consumed in the both of early fusion stages for H-2 and He-3.

**11.4. Mechanism for Producing TSLs in Stellar Nuclear Fusion**

Two stellar nuclear fusion stages consume TSLs, as already described in section 11.1 and 11.2 (Figure 25 and 27). This section explains the mechanism for producing TSLs: It comes from two sources:
- Released from the fusion stage to build He-4 (see Figure 28), and
- Generated from the beginning fusion stage to build H-2 – this will be described in following.

The generation mechanism is composed of two nuclear transformation parts inside the fusion process H-1 + H-1 ➔ H-2 + $e^+$ + υ:

- ✓ Virtual inverse beta decay (Figure 19): $\bar{\nu}_e + p \rightarrow e^+ + n$
- ✓ Electron-proton annihilation (Figure 20): $e^+ + e^- $ ➔ $2\gamma$ + **TSL**

What happens is only possible to consider a pair of hydrogen atoms (two hydrogen kernels and two electrons) in a high-density state (Ref. [6]):

**2 H-1** + **2 e⁻**, ionized & energized
➔ $\bar{\nu}_e$+ ( $p^+ + e^-$ ) + ( $p^+ + e^-$ ) ➔ ( $e^+ + n + e^-$ ) + ( $p^+ + e^-$ )
➔ ( $2\gamma$ + **TSL** + n ) + ( $p^+ + e^-$ ) ➔ $2\gamma$ + ( **TSL** + n + $p^+$) + $e^-$
➔ $2\gamma$ + **H-2** + $e^-$

With the considered group of two hydrogen atoms (already ionized), one up quark of the first H-1 is transformed to a neutron, producing one positron which is soon annihilated together with the first electron, producing gamma photons and a TSL. By consuming this TSL, the second H-1 merges with the transformed neutron to a deuterium kernel H-2 (Figure 9). The second electron balances this merged kernel electrically. The standard description (H-1 + H-1 ➔ H-2 + e⁺ + υ) gets here a more precise and detailed formulation.

The self-providing mechanism for TSL will also take effect for the second stage: H-1 + H-2 ➔ He-3 + γ, and Figure 27, concerning the participating hydrogen atom H-1:

$$\bar{\nu}_e + (p^+ + e^-) + \ldots \to (e^+ + n + e^-) + \ldots \to (2\gamma + \text{TSL} + n) + \ldots$$

Lastly, the third stage of fusion process for building stable helium He-4 releases a TSL, which can participate other fusion stages.

TSL emergency and its involvement in the nuclear fusion processes is new and important in comparison to the standard model. TSL is not easily detectable, thus out of focus of our experimental physics until now. It is a stable state of matter occurrence.

## 12. Essentials of a Nuclear Fusion Process

A very essential thing happened in association with a nuclear fission process will be considered in this section: How a separation occurs at the fraction place, based on the YY model.

Consider the example nuclear fission of uranium atoms (Ref. [3]):
    n + U-92/235 ➔ Ba-56/139 + Kr-36/95 + 2 n + 200 MeV

Without modelling the complete big atom kernel of uranium 235 (that would massively extend the scope of this paper), the single place is interested which separates the uranium atom kernel into the both parts of Ba-139 and Kr-95.

The YY model will describe this place with a possible structure configuration before and after separation in Figure 29. Colored parts help to identify them in the structure of big kernel before and kernel pieces after the separation.

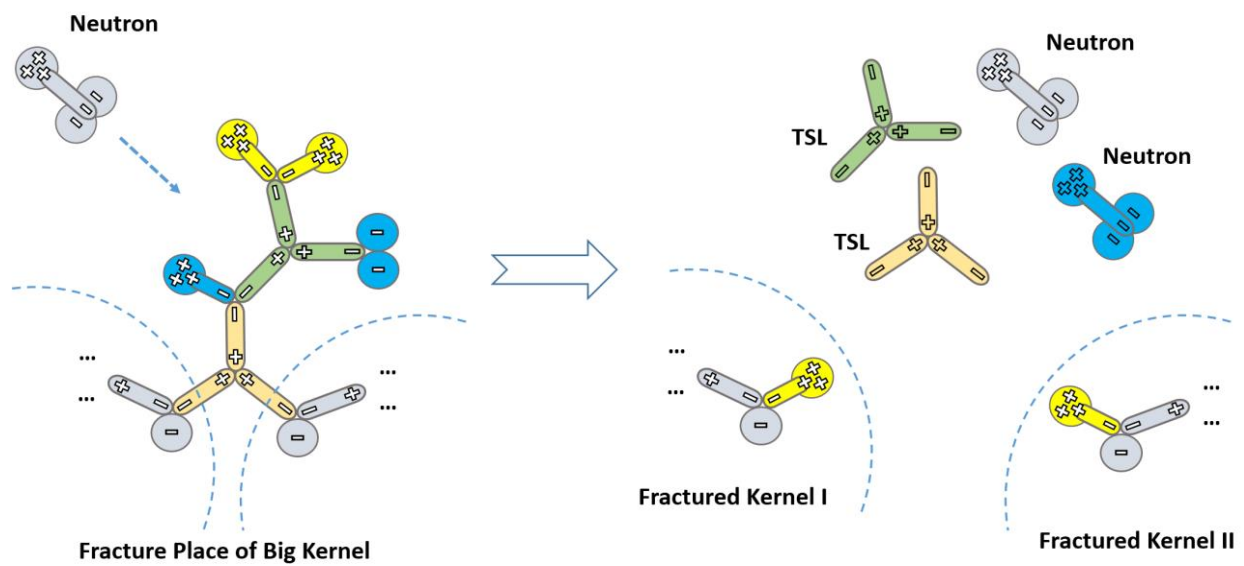

Figure 29: Fission a Big Atom Kernel and its Products

Not only up and down quarks are preserved, but the electrical charges before and after the separation. Additionally, two existing triple space links TSLs are released.

Allowedly, the model description above supposes, that there is a single place to separate the U-235 atom kernel to its fission products Ba-139 and Kr-95. It is possible the separation occurs cross multiple connected places – the final answers will turn out after modelling the complete uranium atoms. Many different valid structure configurations are possible, as the simple cases for tritium H-3 (Figure 11), and for helium He-4 (Figure 13, 16 and 17) already illustrated.

**Cold Fusion:**

YY model provides additional arguments they may help to understand if the cold fusion (Ref. [8]) may work.

A nuclear reaction without consuming TSL may be considered as "cold", because no electron-positron annihilation occurs, thus no gamma radiations are emerged. Transformation processes, that even release TSLs, become "cool" – they offer TSLs for fusions consuming TSLs, without themselves emerging gamma radiations.

Stellar Fusion processes that need TSLs for binding, as already described for building H-2 and He-3, must "self-produce" TSLs and gamma photons.

Because the nuclear fission (n + U-92/235 → Ba-56/139 + Kr-36/95 + 2 n +…) also releases TSLs, it is possible that other big atomic kernels (not necessarily uranium) also do this so that these TSLs can be consumed by parallel running fusions without (or with few) gamma radiations.

Following Figure 30 is the YY model description for well-known nuclear fusion process for helium He-4 (Ref. [6]): H-2 + H-3 → He-4 + 3.5 MeV + (n + 14.1 MeV)

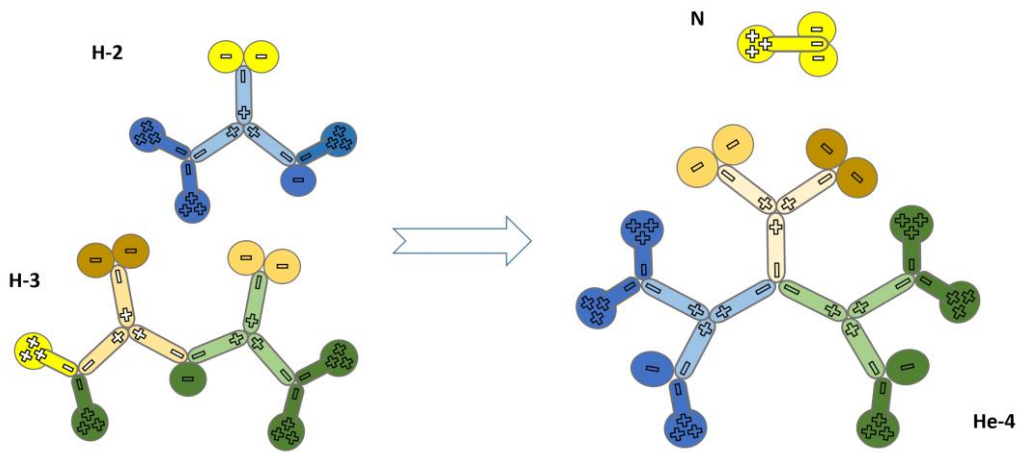

Figure 30: Fusion Process H-2 + H-3 → He-4 + …

No TSL is involved in this process, neither consumed nor released. Gamma radiations (e⁻ + e⁺ → TSL + 2γ) do not occur. Thus, hydrogen fusion from H-2 and H-3 to helium He-4 becomes a good technical candidate for nuclear cold fusion, being interesting for human applications.

## 13. Space-Matter Evolution during Cosmological Big Bang Inflation

Based on the YY model, it is possible to Figure out the major stages of space-matter evolution at some points of the big bang phase (Ref. [9]), especially the state transformations into each other.

### 13.1. Homogeneous Space Structure with High Energy Density

At a very early stage, during the inflation phase after the big bang, the space processes a homogeneous structure, caused by a very high energy density. This structure here (called "**Triple Space Linked State**" - **TSLS**") is also the base for the further matter-building, Figure 31.

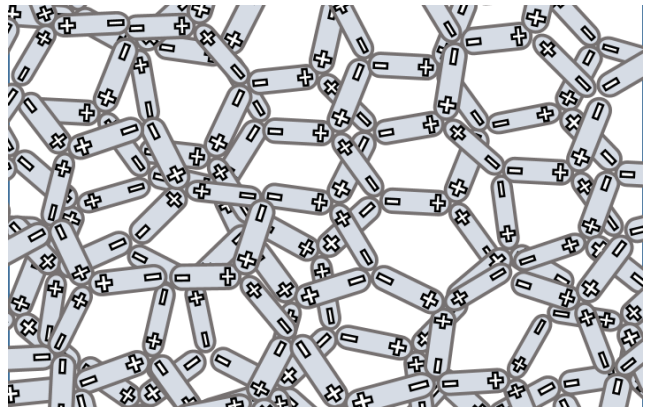

Figure 31: Triple Space Linked State

It is mainly a space structure based on electrostatic interactions of TSLs – The cosmos is filled with Y particles. Matter-antimatter asymmetry of the cosmos can already be founded here. The space is "networked" at this stage of cosmological expansion by inter-linked TSLs.

### 13.2. Pre-build Matter in Form of Up and Down Quarks from TSLs

In this stage, inter-linked TSLs will be released from each other. Up quarks and down quarks will be created from TSLs as manifestation of energy to matter: Every TSL itself, forced by energy, can be "transformed" into one up quark and two down quarks – cutting two yin's and fill all parts with matter, see Figure 32.

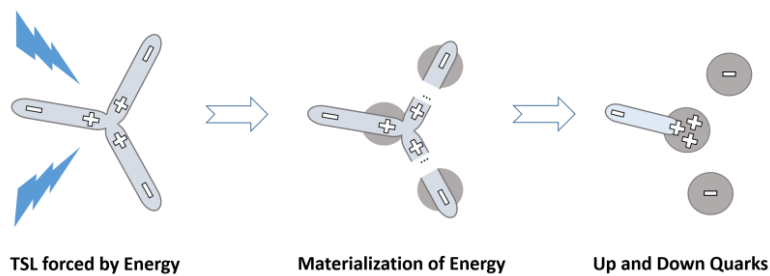

Figure 32: From one TSL to one Up Quark and Two Down Quarks

Because up and down quarks are dependent matter form, "pre-build" of matter is used here. At this stage, space is filled with up quarks, twice times numbers more with down quarks and rest TSLs, Figure 33.

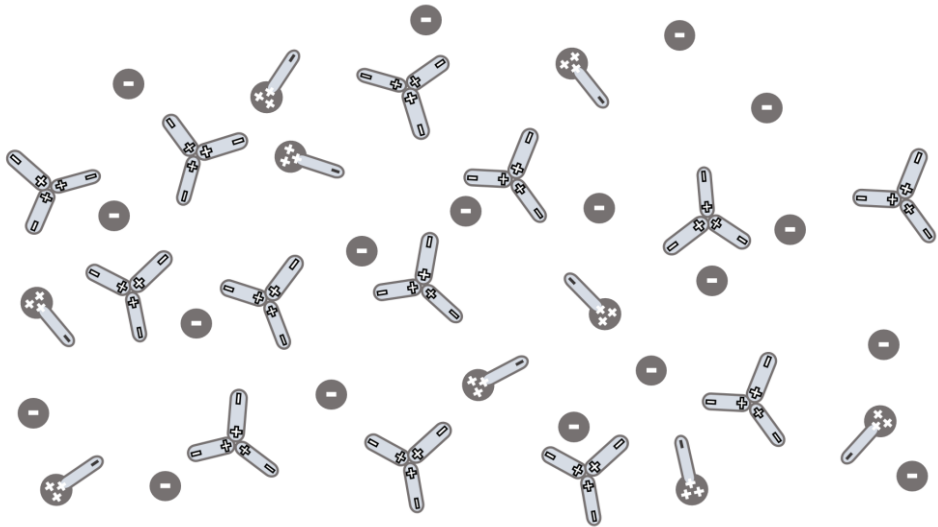

Figure 33: Universe filled with Up Quarks, Down Quarks and TSLs

### 13.3. Build Neutrons

The following stage builds neutrons, whereas the space is expanding and more transparency for particles and more freedom for pre-build matters given. One neutron is assembled by "tripling" one up and two down quarks, Figure 34.

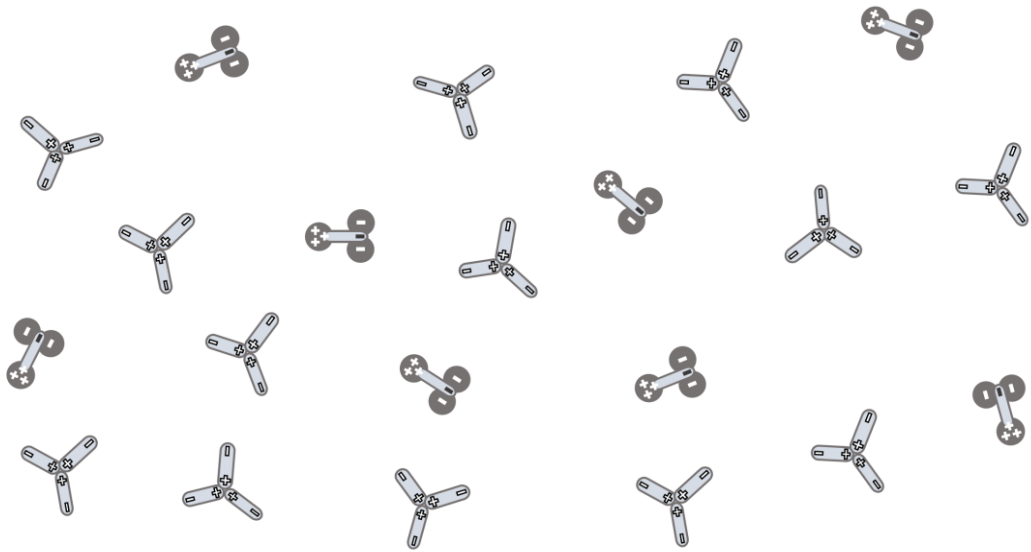

Figure 34: Universe filled with Neutrons and TSLs

Building neutrons can be regarded as one scenario. It is also possible that protons are built with existing up and down quarks parallel to neutrons.

### 13.4. Build Protons and Electrons

As already described in the Figure 18 of section 8, pairs of neutrons can reassemble themselves to protons and electrons. The universe filled with neutrons (Figure 34) will than descend soon into a universe filled partially with protons, electrons and rest neutrons (Figure 35):

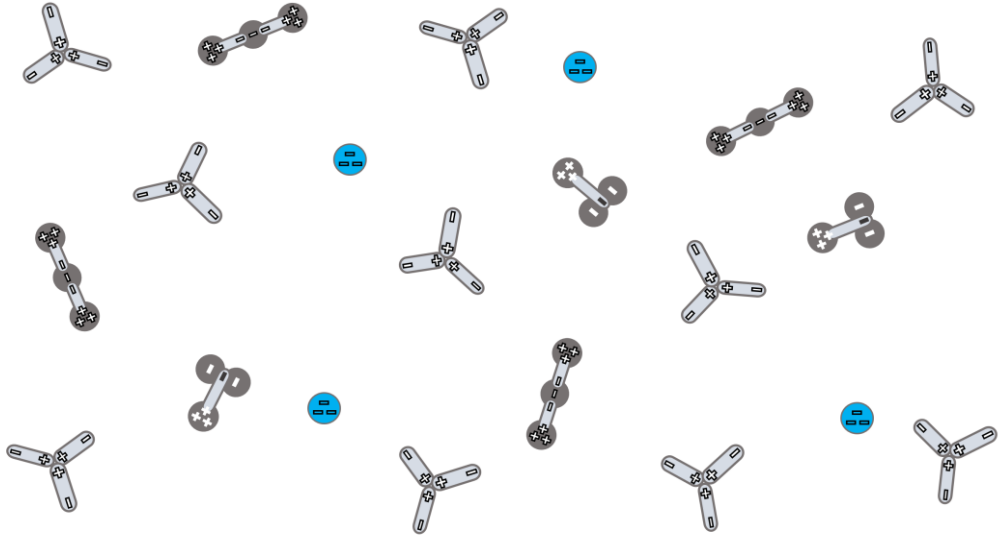

Figure 35:  Universe filled with Protons, Electrons and Neutrons

Neutrinos and antineutrinos are emerged (not drawn in the Figure) by electron prototypes as mediator, see section 8.

### 13.5.  Pairing of Protons and Electrons to Hydrogen Atoms H-1

With expansion of space, free electrons are bound together with protons to form hydrogen atoms (Figure 36). Later hydrogen atoms are grouped to hydrogen molecules by chemical binding.

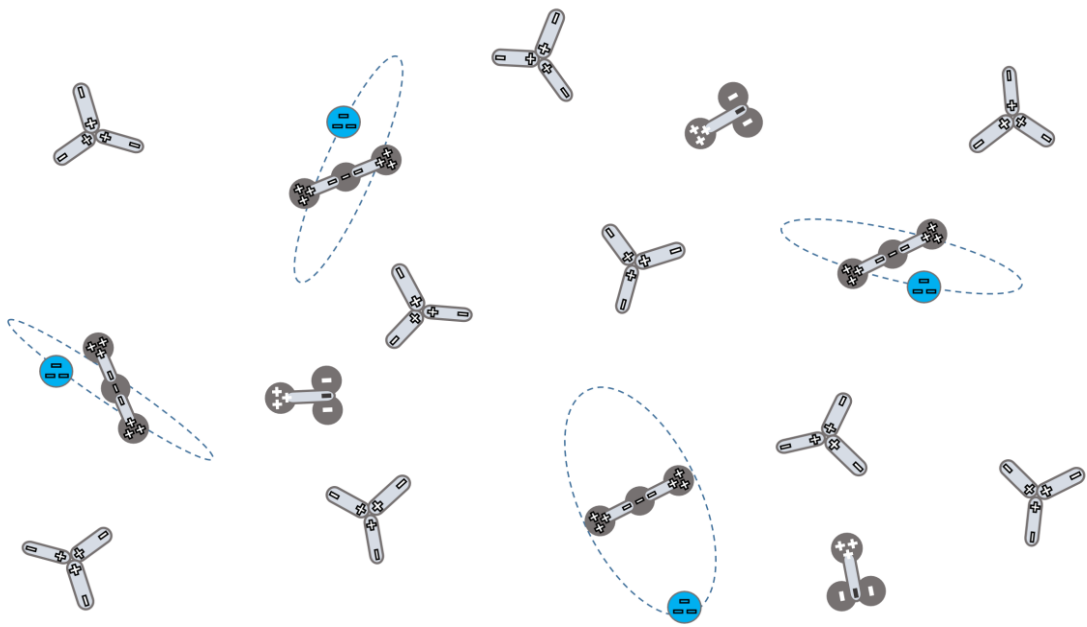

Figure 36:  Universe filled with Hydrogen Atoms and rest Neutrons and TSLs

Two essential details need to be researched: Firstly, how the cosmological inflation leads to a state "filled with" triple space links (TSLs) by considering how high energy interactions go on, while space is expanding at a very early stage of inflation. Secondly, how energies are manifested to pre-build matters (up and down quarks) in the subsequent process of expansion.

## 14. Dark Matter Consideration

Dark matter postulated from cosmological research is still outside the experimental range of our detection technologies. TSLs in the YY model are candidates for dark matter, or they have something strong to do with dark matter.

Triple Space Links TSLs, as a product of the release of Triple Space Linked State (TSLS, Section 13.1), somehow form "low-state" or "reaction-rotten" matter. They are said to have a small mass unit and, more importantly, to be charge-neutral. They are the initial structure for the construction of matter when high energies are applied to them. They are also the product of an electron-positron annihilation, as described in section 8, accompanied by gamma radiation. As shown in earlier sections, they are released during some nuclear fusion and fission processes (Figures 28 and 29). They can be consumed by other fusion processes (Figures 25 and 27).

Obviously, the YY model can explain why TSLs are lost outside our current observation and detection technologies. Since it is possible to model their occurrences and consumption during the nuclear transformation and annihilation processes, and since their role in the construction of the nucleus and their number are known, it should be possible to find some experimental methods to calculate their "dark mass". They could verify the physical existence of TSLs – the Y-particle - by balancing all measurable masses and energies and postulated dark mass participations.

## 15. Outlook

The YY model opens up a broad spectrum of theoretical and experimental research in atomic physics and molecular chemistry. On the basis of the detailed construction structure of the atomic nucleus and its variants, electrostatic and quantum field calculations could deliver more filigree results. For example, the three-dimensional structure of a heavy atomic nucleus can be contracted in some parts and stretched or folded in other parts only by electrostatic interactions between protonheads and neutronheads (both negatively charged) and protonides (negatively charged).

Some very important aspects in the YY model are the handling of PSL and TSL. It requires a mathematical model that must include considerations of energy and matter to explain how these connections work. So far they are assumptions in the YY model.

The distribution of electron clouds in orbitals around the atomic nucleus will also be important. A more detailed atomic model could reveal more details at the molecular level that have been hidden until now.

An isotope of an atom can have different kernel structure configurations, as already shown for tritium H-3 (Figure 11) and for helium He-4 (Figures 13, 16 and 17). Research for differences in physical and chemical properties can reveal interesting aspects.

**References**

- 22 -

**Authors:**

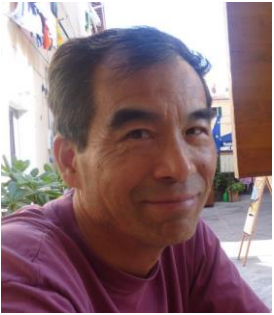

*Dr Ing Hongguang Yang (hyang2013@gmail.com) studied Astronomical and Physical Geodesy and Physical Sciences at the Technical University of Munich there he also made his PhD in Geodesy. He is currently working in the area of computer sciences. Born in China, he is living in Germany.*

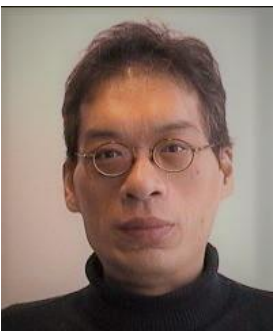

*Dr ret nat Weidong Yang (wdyconf@gmail.com) studied Physical Sciences at the University of Augsburg there he also made his PhD in Solid State Physics. He is currently working in the area of computer sciences. Born in China, he is living in Germany.*